\documentclass[twocolumn]{jpsj3}

\usepackage{bm}
\usepackage{color}
\usepackage{afterpage}
\usepackage{multicol}

\def\runauthor{Online-Journal Subcommittee}

\title{%
Construction of a Neural Network with Temperature-Dependent Recall Patterns\\
}

\author{
Munetaka \textsc{Sasaki}\thanks{E-mail : msasaki@kanagawa-u.ac.jp}}

\inst{
Faculty of Engineering, Kanagawa University, Yokohama 221-8686, Japan
}

\recdate{\today}

\abst{
We present a simple model that recalls two different patterns depending on the temperature. 
To realize a change in recall pattern due to temperature change, we embed two patterns to different graphs: 
the first pattern into a fully connected graph and the second pattern into a sparse graph. 
Because a fully connected graph is more resistant to thermal fluctuations than a sparse graph, 
we can realize a change in recall pattern by tuning relative weights of the two patterns properly. 
We demonstrate by equilibrium Monte-Carlo simulations that such a temperature-dependent change in recall patterns does occur in our model. 
Simulation results strongly indicate that the system undergoes a first-order phase transition when the change in recall patterns occurs. 
It is also demonstrated by annealing simulations that the system fails to recall the pattern embedded in the sparse graph at low temperatures 
if the free-energy barrier is too high to overcome within the given simulation timescale. 
}

\begin{document}
\maketitle

\section{Introduction}
The Hopfield model~\cite{Hopfield82} has served as a paradigmatic model of associative memory in neural systems. 
The Hamiltonian of the Hopfield model is defined by
\begin{equation}
{\cal H}=-\sum_{i<j} J_{ij}S_i S_j,
\label{eqn:Hopfield}
\end{equation}
where $S_i$ is an Ising variable whose value is either $+1$ or $-1$. 
The state of neuron $i$ is described by this variable. 
The coupling constant $J_{ij}$ is given by the Hebb rule~\cite{Hopfield82, Hebb49}
\begin{equation}
J_{ij}= \frac{1}{N}\sum_{\mu=1}^{P} \xi_i^\mu \xi_j^\mu,
\label{eqn:Hebb_Rule}
\end{equation}
where $N$ is the number of neurons, $\{\xi_i^{\mu} \}$ is the $\mu$-th pattern stored to the model, 
and $P$ is the number of patterns. We find from these equations that in the Hopfield model, 
all patterns are embedded in a fully connected graph because the sum in Eq.~\eqref{eqn:Hopfield} 
is taken over all pairs and $J_{ij}$ defined by Eq.~\eqref{eqn:Hebb_Rule} is non-zero for all pairs with $i\ne j$. 
We also notice from Eq.~\eqref{eqn:Hebb_Rule} that the same weight is assigned to all patterns. 

In this study, we propose a simple model that recalls two different patterns depending on the temperature. 
To construct such a neural network, we embed two patterns to different graphs: the first pattern into a fully connected graph 
and the second pattern into a sparse graph. The key point is that mean-field-like cooperative systems defined on 
fully connected graphs are more resistant to thermal fluctuations than systems defined on sparse graphs. 
Therefore, by tuning relative weights of the two patterns properly, we can construct a neural network that recalls the pattern 
embedded in the fully connected graph at high temperatures and that embedded in the sparse graph at low temperatures. 
In this study, we choose the two dimensional square lattice as a sparse graph. 
We demonstrate by equilibrium Monte-Carlo simulations that such a temperature-dependent change in recall patterns does occur in our model. 
Simulation results strongly indicate that the system undergoes a first-order phase transition when the change in recall patterns occurs. 
We also performed an annealing simulation in which the temperature was gradually decreased from high to low. 
Consequently, we found that the system fails to recall the pattern embedded in the sparse graph at low temperatures 
if the free-energy barrier is too high to overcome within the given simulation timescale. 

The outline of the paper is as follows: In Sect.~\ref{sec:Model}, we define our model. 
In Sect.~\ref{sec:Method}, we describe our simulation method. 
In Sect.~\ref{sec:Results}, we show our simulation results. 
Section~\ref{sec:Summary_Discussion} is devoted to summary and discussion.

\section{Model}
\label{sec:Model}
To construct a neural network that recalls two different patterns depending on the temperature, 
we consider the following Hamiltonian with two terms:
\begin{equation}
{\cal H}={\cal H}^{\rm S}+{\cal H}^{\rm F}.
\label{eqn:Hamiltonian}
\end{equation}
The first and the second terms represent contribution from a sparse graph and that 
from a fully connected graph, respectively. The superscript "S" denotes the sparse graph, and "F" denotes 
the fully connected graph. 

The first term ${\cal H}^{\rm S}$ in Eq.~\eqref{eqn:Hamiltonian} is given as
\begin{equation}
{\cal H}^{\rm S}\{S_i\}=-\sum_{\langle ij \rangle} J_{ij}^{\rm S} S_i S_j, 
\label{eqn:Hamiltonian_S}
\end{equation}
where $S_i$ is an Ising variable located on site $i$. The value of $S_i$ is either $-1$ or $+1$. 
The sum on the right-hand side runs over all two sites that are connected by an edge in the sparse graph. 
The coupling constant $J_{ij}^{\rm S}$ is given as
 \begin{equation}
J_{ij}^{\rm S}=J\xi_{i}^{\rm S} \xi_{j}^{\rm S},
\label{eqn:CouplingConstant_S}
\end{equation}
where $J(>0)$ is an exchange constant and $\{\xi_i^{\rm S}\}$ denotes a pattern embedded 
in the sparse graph. The possible value for $\xi_i^{\rm S}$ is either $-1$ or $+1$. 
It is worth pointing out that ${\cal H}^{\rm S}$ is a Hamiltonian of the Mattis model~\cite{Mattis76} defined 
on the sparse graph. Therefore, the system with the Hamiltonian ${\cal H}^{\rm S}$ is 
equivalent to the ferromagnetic Ising model on the sparse graph. 
The ground state of ${\cal H}^{\rm S}$ is $\pm \{\xi_i^{\rm S}\}$. 

The second term ${\cal H}^{\rm F}$ in Eq.~\eqref{eqn:Hamiltonian} is given as
\begin{equation}
{\cal H}^{\rm F}\{S_i\}=-\frac{kNC^{\rm F}J}{2}\left(\bar{m}^{\rm F}\{S_i\}\right)^2,
\label{eqn:Hamiltonian_F}
\end{equation}
where $N$ is the number of sites, $C^{\rm F} (>0)$ is a parameter that adjusts a balance between 
${\cal H}^{\rm S}$ and ${\cal H}^{\rm F}$, and $k$ is the degree of the sparse graph. 
We assume that all the vertices have the same degree $k$ in the sparse graph. 
$\bar{m}^{\rm F}$ is a "magnetization" of the fully connected graph defined by
\begin{equation}
\bar{m}^{\rm F}\{S_i\} \equiv \left|\frac{1}{N}\sum_i \xi_i^{\rm F}S_i\right|,
\label{eqn:magnetization_F}
\end{equation}
where $\{\xi_i^{\rm F}\}$ denotes a pattern embedded in the fully connected graph. 
Again, the possible value for $\xi_i^{\rm F}$ is either $-1$ or $+1$. 
For simplicity, we assume that $\xi_i^{\rm S}$ and $\xi_i^{\rm F}$ 
are randomly and independently determined either $+1$ or $-1$ with probability $1/2$ for all $i$. 
The Hamiltonian ${\cal H}^{\rm F}$ is equivalent to that of the Mattis model 
defined on the fully connected graph because we can rewrite ${\cal H}^{\rm F}$ as
\begin{equation}
{\cal H}^{\rm F}\{S_i\}=-\frac{kC^{\rm F}}{N}\sum_{i<j}J_{ij}^{\rm F} S_i S_j-\frac{kC^{\rm F}J}{2}, 
\label{eqn:Hamiltonian_F_v2}
\end{equation}
where $J_{ij}^{\rm F}\equiv J\xi_{i}^{\rm F} \xi_{j}^{\rm F}$. The transition temperature 
$T_{\rm c}^{\rm F}$ of ${\cal H}^{\rm F}$, which is equal to that in the infinite-range 
ferromagnetic Ising model, is given as
\begin{equation}
T_{\rm c}^{\rm F}=kC^{\rm F}J,
\label{eqn:Tc_F}
\end{equation}
where we set the Boltzmann constant $k_{\rm B}$ to unity. 

Now the point is that we can adjust the the transition temperature $T_{\rm c}^{\rm S}$ of ${\cal H}^{\rm F}$ 
by changing the sparse graph. For example, if we choose the two dimensional square lattice 
as a sparse graph, the transition temperature $T_{\rm c}^{\rm S}$ becomes $2/\log(1+\sqrt{2})J\approx 2.27J$. 
The degree $k$ of the two dimensional square lattice is four. Therefore, 
if $C^{\rm F} > T_{\rm c}^{\rm S}/(4J) \approx 0.567$, 
$T_{\rm c}^{\rm F}$ becomes larger than $T_{\rm c}^{\rm S}$. On the other hand, 
the ground state energy of ${\cal H}^{\rm S}$ and that of ${\cal H}^{\rm F}$, 
which are denoted as $E_{\rm GS}^{\rm S}$ and $E_{\rm GS}^{\rm F}$ respectively, 
are given as
\begin{equation}
E_{\rm GS}^{\rm S} =-\frac{kNJ}{2},\quad E_{\rm GS}^{\rm F} =-\frac{kNC^{\rm F}J}{2}.
\label{eqn:E_GS}
\end{equation}
Therefore, if $C^{\rm F}<1$, $E_{\rm GS}^{\rm S}$ is smaller than $E_{\rm GS}^{\rm F}$. 

Now let us consider what happens when $0.567 < C^{\rm F} < 1$. As the temperature is decreased from a 
high temperature, the system with the Hamiltonian ${\cal H}$ of Eq.~\eqref{eqn:Hamiltonian} 
is expected to undergo a phase transition at $T_{\rm c}^{\rm F}$ because $T_{\rm c}^{\rm F} > T_{\rm c}^{\rm S}$. 
When $T_{\rm c}^{\rm S} < T < T_{\rm c}^{\rm F}$, the order parameter $\bar{m}^{\rm F}$ defined by Eq.~\eqref{eqn:magnetization_F} grows with decreasing the temperature. However, we can expect that 
a second transition occurs at $T^{\rm S}$ and the order parameter $\bar{m}^{\rm S}$ defined by 
\begin{equation}
\bar{m}^{\rm S}\{S_i\} \equiv \left|\frac{1}{N}\sum_i \xi_i^{\rm S}S_i\right|,
\label{eqn:magnetization_S}
\end{equation}
grows as the temperature is further decreased because $E_{\rm GS}^{\rm S} < E_{\rm GS}^{\rm F}$. 
As shown in appendix, we can always choose $C^{\rm F}$ for any sparse graphs 
so that such sequential transitions occur. 

\section{Method}
\label{sec:Method}
In this study, we use a variant of the Wang-Landau method regarding the order parameter~\cite{WatanabeSasaki11}. 
In this method, the Wang-Landau method~\cite{WangLandau01A, WangLandau01B} is modified 
to measure free energy as a function of order parameters. A similar method has been proposed 
in Ref.~\citen{Berg93} to calculate free energy by the multicanonical ensemble 
method~\cite{BergNeuhaus91, BergNeuhaus92}. In the present study, we measure free energy 
as a function of order parameters $m^{\rm S}$ and $m^{\rm F}$ defined by
\begin{eqnarray}
&&\exp[-\beta F(\beta;m^{\rm S},m^{\rm F})] \nonumber \\
&&\equiv C {\rm Tr}_{\{S_i\}} \exp[-\beta {\cal H}\{S_i\}]\delta(m^{\rm S}-\bar{m}^{\rm S}\{S_i\}) \nonumber \\
&&\quad \times \delta(m^{\rm F}-\bar{m}^{\rm F}\{S_i\}),
\label{eqn:FeneDef}
\end{eqnarray}
where $C$ is a constant and $\beta$ is the inverse temperature. It is not important how we choose $C$ 
because it only contributes to $F(\beta;m^{\rm S},m^{\rm F})$ as a constant. It is worth pointing out that 
$\exp[-\beta F(\beta;m^{\rm S},m^{\rm F})]$ is proportional to the probability $P(\beta;m^{\rm S},m^{\rm F})$ that a state with magnetizations $(m^{\rm S},m^{\rm F})$ is sampled at temperature $T$. Therefore, 
by measuring the free energy $F(\beta;m^{\rm S},m^{\rm F})$, we can easily estimate the thermal averages of two magnetizations as 
\begin{eqnarray}
&&\hspace*{-7mm}\langle m^{\rm S} \rangle_T =\frac{1}{Z(\beta)}\iint{\rm d}m^{\rm S}{\rm d}m^{\rm F} m^{\rm S}
\exp[-\beta F(\beta;m^{\rm S},m^{\rm F})], \nonumber \\
&&\hspace*{-7mm}\langle m^{\rm F} \rangle_T =\frac{1}{Z(\beta)}\iint{\rm d}m^{\rm S}{\rm d}m^{\rm F} m^{\rm F}
\exp[-\beta F(\beta;m^{\rm S},m^{\rm F})], \nonumber \\
\label{eqn:MaveDef}
\end{eqnarray}
where
\begin{eqnarray}
Z(\beta)\equiv \iint{\rm d}m^{\rm S}{\rm d}m^{\rm F} \exp[-\beta F(\beta;m^{\rm S},m^{\rm F})].
\end{eqnarray}

We next explain the detailed conditions of simulations. In the present study, we chose 
an $L\times L$ square lattice with periodic boundary conditions as a sparse graph. 
Therefore, the degree $k$ is four. The values of $L$ we investigated are $10,~20,$ and $30$. 
The number of sites $N$ is $L^2$. 
We hereafter use $J$ as a unit of temperature. 
For each $L$, we performed free-energy measurement for $100$ different samples with different 
embedded patterns $\{\xi_i^{\rm S}\}$ and $\{\xi_i^{\rm F}\}$ at temperatures ranging from 
$1.0~J$ to $6.0~J$. In the free-energy measurement, we set the initial value of 
the modification constant $\Delta F$ in Ref.~\citen{WatanabeSasaki11}, 
which is related with the modification factor $f$ in the Wang-Landau method by $f=\exp(\Delta F)$, 
to unity. We stopped our simulation after we halved $\Delta F$ 
$20$ times. Therefore, the final $\Delta F$ is $2^{-20}$. 
The histogram of two magnetizations $H(m^{\rm S},m^{\rm F})$ was checked every $10,000$ 
Monte-Carlo steps (MCS). We regarded the histogram as flat when $H(m^{\rm S},m^{\rm F})$'s for all the magnetizations 
are not less than $80\%$ of the average value of the histogram. 

\begin{figure}[h]
\begin{center}
\includegraphics[width=9cm]{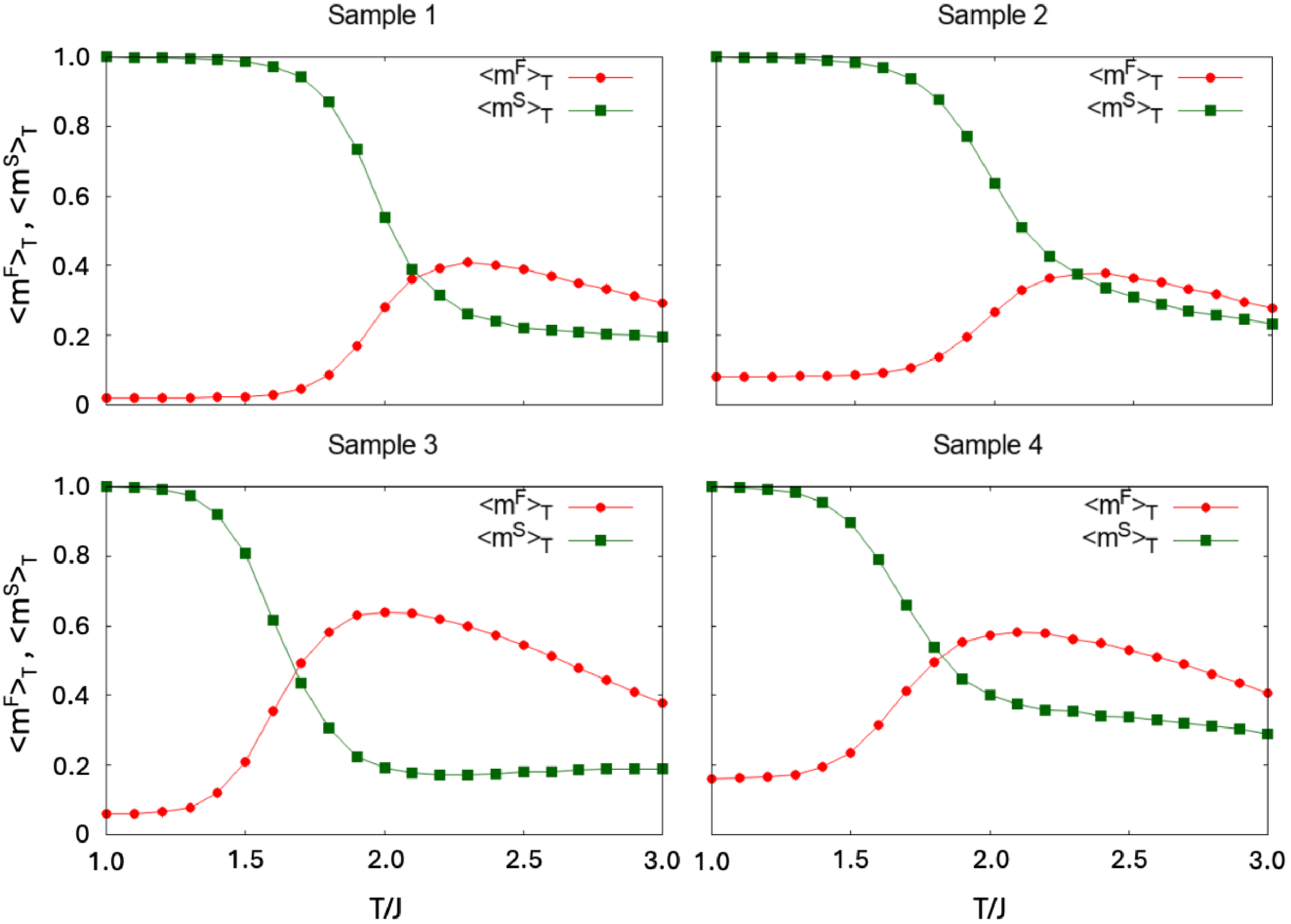}
\end{center}
\caption{The temperature dependences of $\langle m_{\rm F} \rangle_T$ and 
$\langle m_{\rm S} \rangle_T$ for four samples. The size $L$ is $10$ and 
the coefficient $C^{\rm F}$ is $0.9$.}
\label{fig:SampMag_L010_CMF090}
\end{figure}

\section{Results}
\label{sec:Results}
\subsection{Magnetization measurement}
\label{subsec:result1}
\begin{figure}[h]
\begin{center}
\includegraphics[width=9cm]{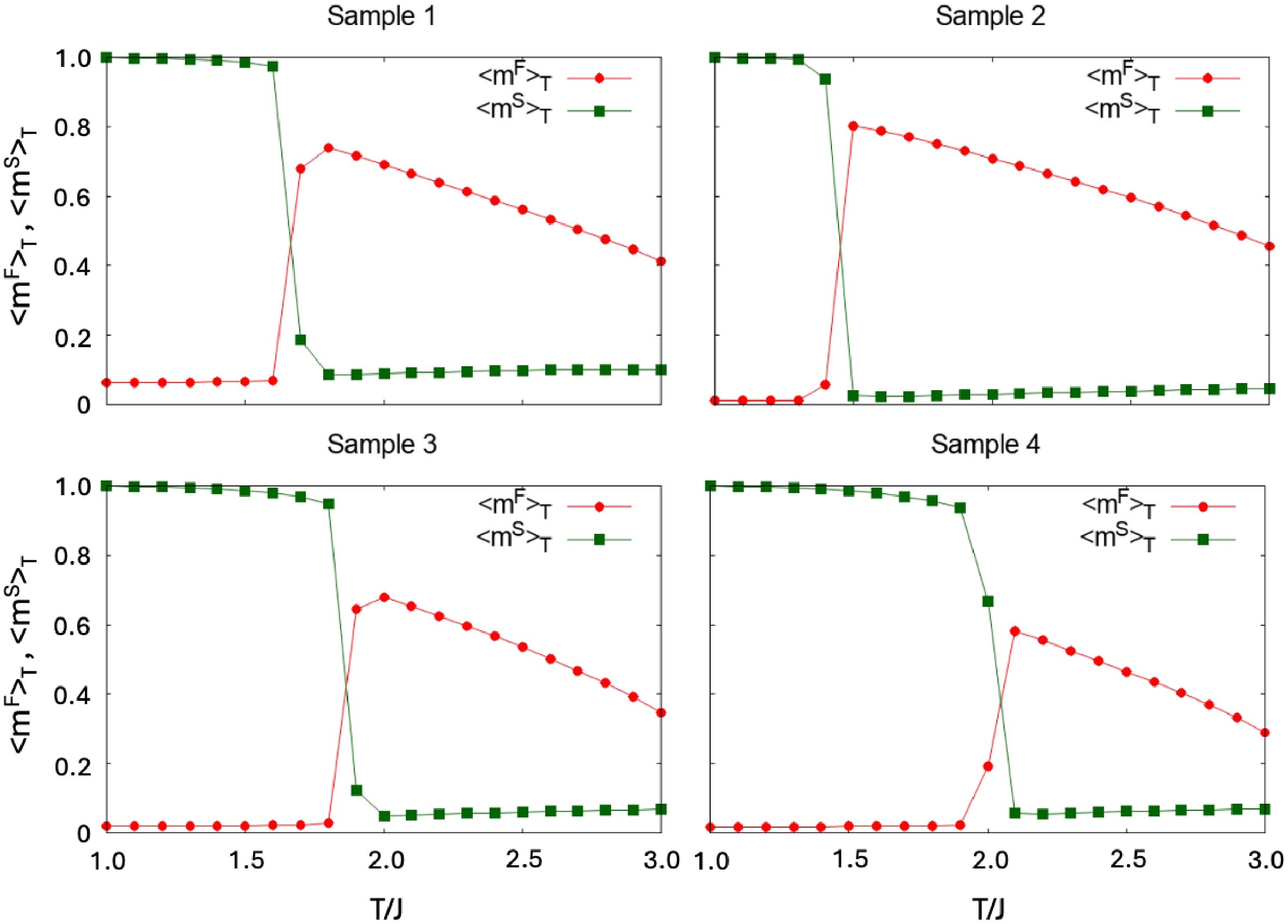}
\end{center}
\caption{The temperature dependences of $\langle m_{\rm F} \rangle_T$ and 
$\langle m_{\rm S} \rangle_T$ for four samples. The size $L$ is $30$ and 
the coefficient $C^{\rm F}$ is $0.9$.}
\label{fig:SampMag_L030_CMF090}
\end{figure}

In Figs.~\ref{fig:SampMag_L010_CMF090} and \ref{fig:SampMag_L030_CMF090}, $\langle m^{\rm S} \rangle_T$ and $\langle m^{\rm F} \rangle_T$ calculated by 
Eq.~(\ref{eqn:MaveDef}) are plotted as a function of $T$ for four samples. The size $L$ 
is $10$ for Fig.~\ref{fig:SampMag_L010_CMF090}  and $30$ for Fig.~\ref{fig:SampMag_L030_CMF090}. 
The coefficient $C^{\rm F}$ is $0.9$ and $T_{\rm c}^{\rm F}$ given by Eq.~\eqref{eqn:Tc_F} is $3.6~J$.
As expected, dominant magnetization changes from $m^{\rm F}$ to $m^{\rm S}$ with decreasing 
temperature. This means that the system recalls the pattern $\{ \xi_i^{\rm F} \}$ embedded in the fully 
connected graph at high temperatures, and the pattern $\{ \xi_i^{\rm S} \}$ embedded in the square lattice 
at low temperatures. Although the switching of the dominant magnetization occurs gradually 
for $L=10$, it occurs abruptly for $L=30$. This result indicates that a first-order phase transition 
occurs with the change in the recalling pattern. We also notice that the crossing temperature $T_{\rm cross}$, 
at which the magnitude relationship between $m^{\rm F}$ and $m^{\rm S}$ changes, is highly 
sample-dependent. 

\begin{figure}[h]
\begin{center}
\includegraphics[width=9cm]{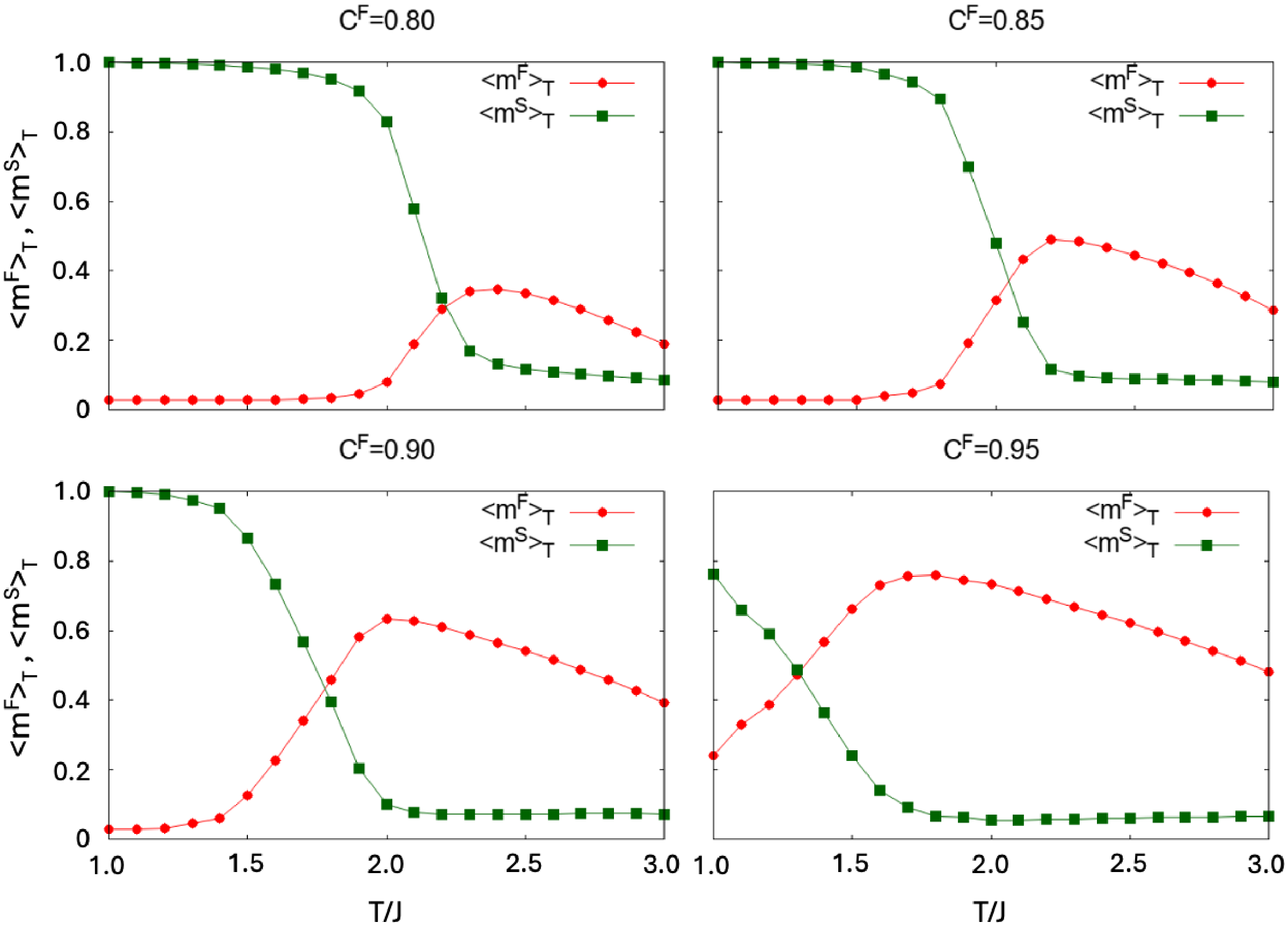}
\end{center}
\caption{The temperature dependences of sample-averaged magnetizations. 
The average was taken over $100$ samples. The size $L$ is $30$. 
The values of $C^{\rm F}$ are $0.80$, $0.85$, $0.90$, and $0.95$ from top left to right bottom.}
\label{fig:AveMag_L030}
\end{figure}

Figure.~\ref{fig:AveMag_L030} shows the temperature dependences of 
sample-averaged magnetizations. The average was taken over $100$ samples. 
The size $L$ is $30$. The values of $C^{\rm F}$ are $0.80$, $0.85$, $0.90$, and $0.95$ 
from top left to right bottom. We see that the change in the dominant magnetization 
occurs gradually because $T_{\rm cross}$ depends on the sample and the average 
is taken over $100$ samples. It should be noted that, as mentioned above, 
the switch in the dominant magnetization in each sample occurs abruptly when $L=30$. 
The switch in the dominant magnetization occurs most clearly when $C^{\rm F}$ is $0.90$. 
Therefore, we hereafter focus on the case of $C^{\rm F}=0.90$. 

\begin{figure}[h]
\begin{center}
\includegraphics[width=7cm]{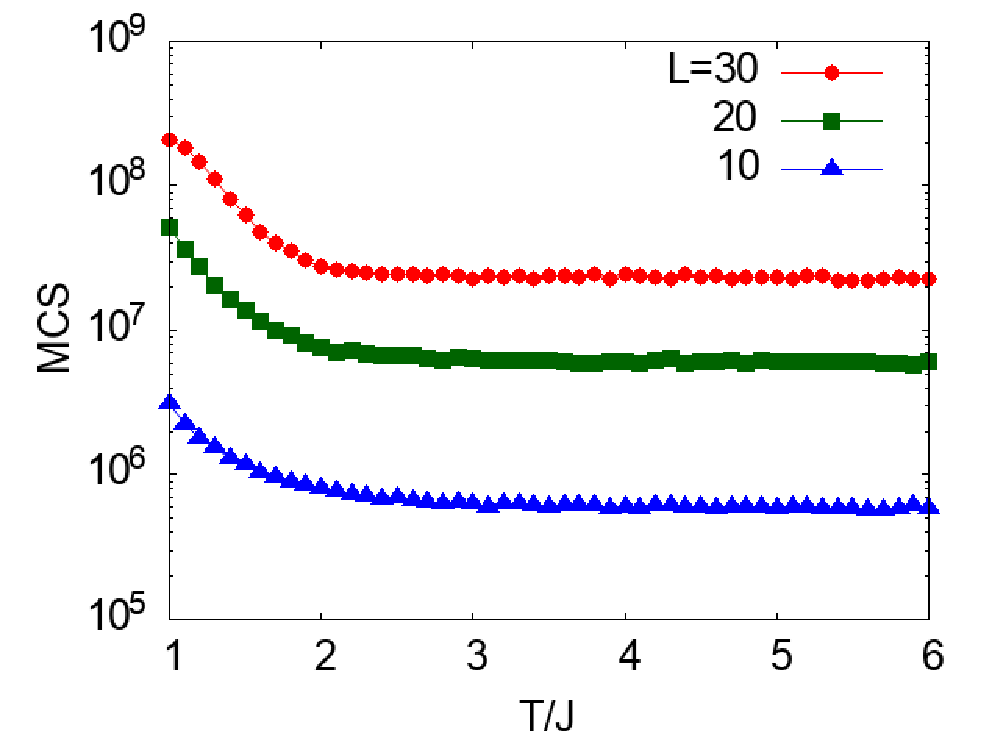}
\end{center}
\caption{The temperature dependence of the total number of MCS required 
to perform a variant of the Wang-Landau method. The coefficient $C^{\rm F}$ is $0.90$. 
The values of $L$ are $10$, $20$, and $30$ from bottom to right. 
The average was taken over $100$ samples.}
\label{fig:MCexit_CMF090}
\end{figure}

In Fig.~\ref{fig:MCexit_CMF090}, the total number of MCS required 
to perform a variant of the Wang-Landau method described in Sect.~\ref{sec:Method} 
is plotted as a function of temperature. The needed MCS increases with increasing $L$. 
We also notice that MCS begins to increase as the temperature is lowered to around $2.0~J$. 
When $L=30$ and $T=1.0~J$, the required MCS is about $2.0\times 10^8$. 
This is the reason why the sizes we investigated in the present study are rather small. 

\subsection{Free-energy analysis}
\label{subsec:result2}
\begin{figure}[h]
\begin{center}
\includegraphics[width=9cm]{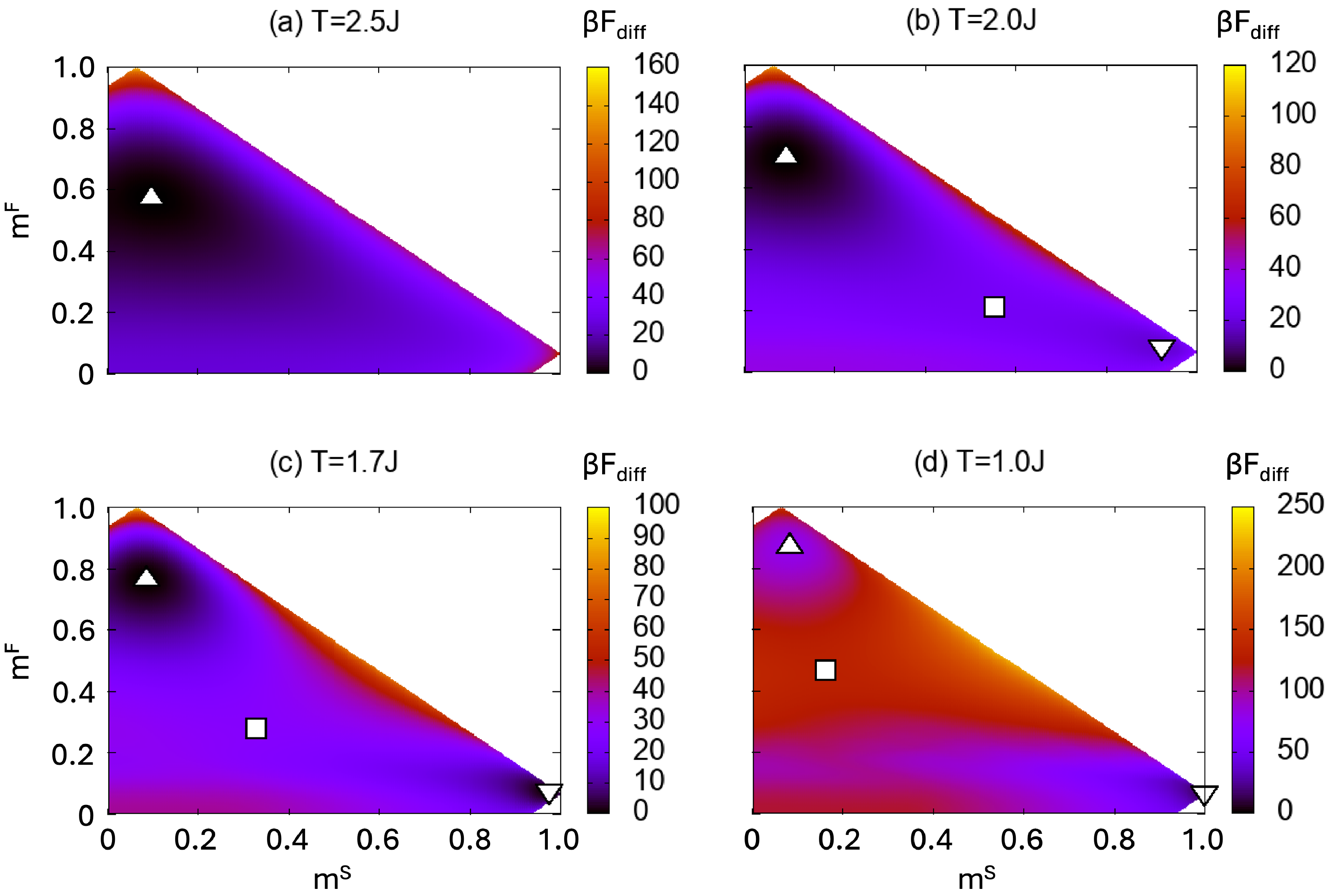}
\end{center}
\caption{$\beta F_{\rm diff}$ of a sample is plotted as a function of two magnetizations for four different temperatures. $F_{\rm diff}$ is defined by Eq.~\eqref{eqn:def_Fdiff}. The sample is 
the same as "Sample 1" in Fig.~\ref{fig:SampMag_L030_CMF090}. 
The size $L$ is $30$ and the coefficient $C^{\rm F}$ is $0.90$. 
The locations of $F_{\rm min}$, $S_{\rm min}$, and $B$ are denoted by triangles, 
inverted triangles, and squares, respectively. See text for the definitions of 
$F_{\rm min}$, $S_{\rm min}$, and $B$.}
\label{fig:Fene_L030_CMF090}
\end{figure}

\begin{figure}[h]
\begin{center}
\includegraphics[width=7cm]{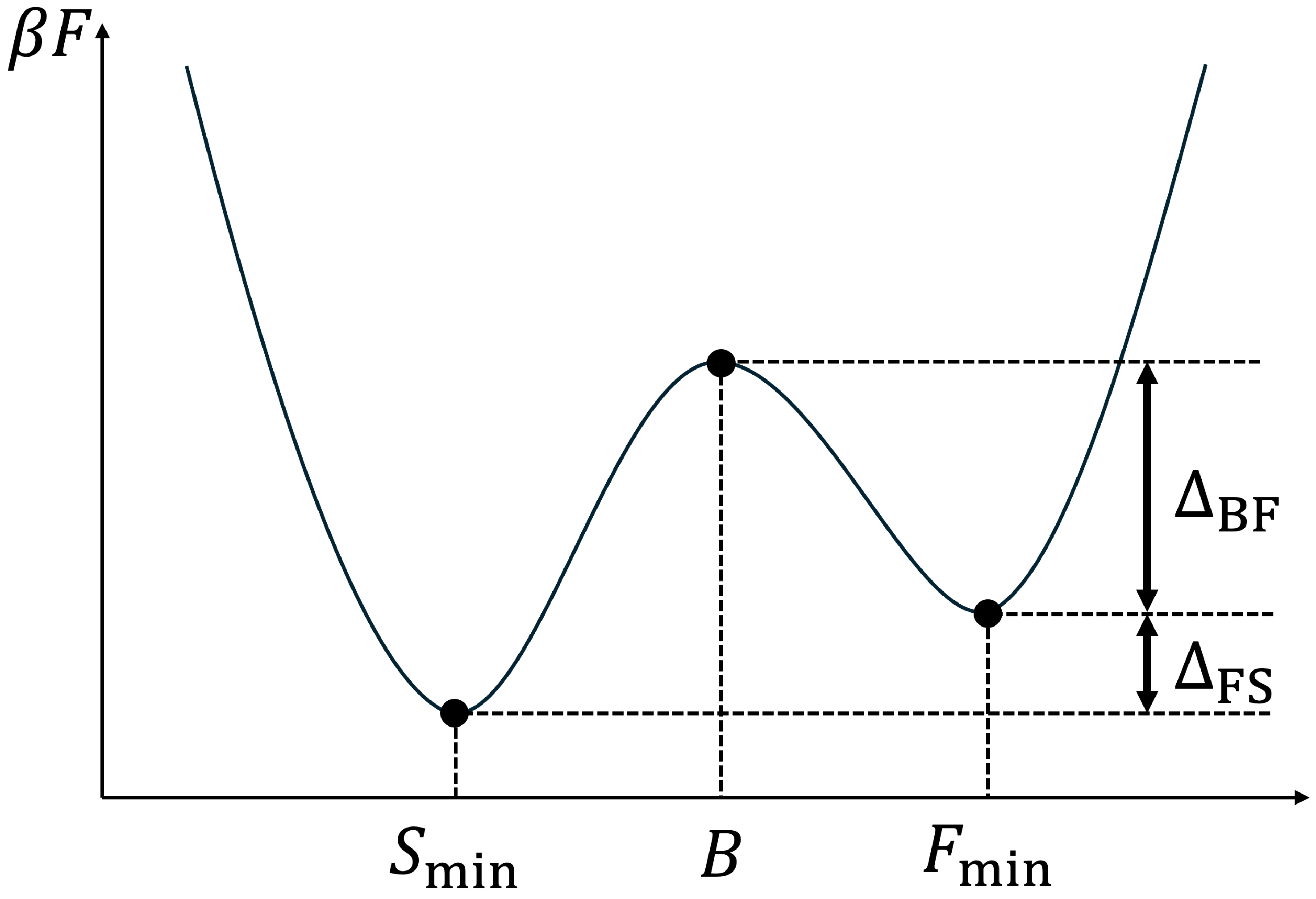}
\end{center}
\caption{The definitions of the free-energy differences $\Delta_{\rm FS}$ and $\Delta_{\rm BF}$.}
\label{fig:Def_DF}
\end{figure}

Figure~\ref{fig:Fene_L030_CMF090} shows a typical result of free-energy measurement of a sample. 
In Fig.~\ref{fig:Fene_L030_CMF090}, $\beta F_{\rm diff}$ is plotted as a function of 
$(m^{\rm S}, m^{\rm F})$, where $F_{\rm diff}$ is the free-energy difference defined by
\begin{align}
F_{\rm diff}(\beta; m^{\rm S}, m^{\rm F})=F(\beta; m^{\rm S}, m^{\rm F})-F_{\rm min}(\beta),
\label{eqn:def_Fdiff}
\end{align}
and
\begin{align}
F_{\rm min}(\beta)=\min\nolimits_{(m^{\rm S}, m^{\rm F})} F(\beta; m^{\rm S}, m^{\rm F}).
\label{eqn:def_Fmin}
\end{align}
The sample is the same as "Sample 1" in Fig.~\ref{fig:SampMag_L030_CMF090}. 
The size $L$ is $30$ and the coefficient $C^{\rm F}$ is $0.90$. 
When $T$ is $2.5~J$, which is between $T_{\rm c}^{\rm S}\approx 2.27~J$ and  $T_{\rm c}^{\rm F}=3.6~J$, 
the free energy has a single local minimum around $(m^{\rm S}, m^{\rm F})\approx (0.10, 0.57)$ 
(Fig.~\ref{fig:Fene_L030_CMF090}~(a)). We hereafter denote a local minimum with $m^{\rm F} > m^{\rm S}$ 
as $F_{\rm min}$. Similarly, a local minimum with $m^{\rm S} > m^{\rm F}$ is denoted as $S_{\rm min}$. 
In Fig.~\ref{fig:Fene_L030_CMF090}, $F_{\rm min}$ and $S_{\rm min}$ are denoted 
by triangles and inverted triangles, respectively. The single local minimum at $T=2.5~J$ is $F_{\rm min}$. 
When $T$ is $2.0~J$, there are two local minima $F_{\rm min}$ and $S_{\rm min}$  (Fig.~\ref{fig:Fene_L030_CMF090}~(b)). 
As schematically shown in Fig.~\ref{fig:Def_DF}, we hereafter denote a barrier that separates the two local minima as $B$. 
In Fig.~\ref{fig:Fene_L030_CMF090}, the barrier $B$ is denoted by squares. 
At this temperature, $F_{\rm min}$ is the global minimum. However, the free energy of $S_{\rm min}$ 
becomes almost equal to that of $F_{\rm min}$ at $T=1.7~J$ (Fig.~\ref{fig:Fene_L030_CMF090}~(c)), 
and $S_{\rm min}$ becomes the global minimum at lower temperatures (Fig.~\ref{fig:Fene_L030_CMF090}~(d)). 
It should be noted that the temperature in Fig.~\ref{fig:Fene_L030_CMF090}~(c), i.e., $1.7~J$, is close to 
the crossing temperature $T_{\rm cross}$ of "Sample 1" in Fig.~\ref{fig:SampMag_L030_CMF090}. 
The location of the global minimum shifts discontinuously as the temperature is changed across $T_{\rm cross}$. 
This behavior explains the reason why two magnetizations in Fig.~\ref{fig:SampMag_L030_CMF090} changes 
abruptly around $T_{\rm cross}$. 
These observations strongly support that, as mentioned above, the transition at $T_{\rm cross}$ is a first-order transition.

We next explain how we analyze these data of free energy. As mentioned before, we first define 
$T_{\rm cross}$ as a temperature at which the magnitude relationship between $m^{\rm F}$ and $m^{\rm S}$ changes. 
We estimate $T_{\rm cross}$ by interpolation. 
We next define the free-energy differences $\Delta_{\rm FS}$ and $\Delta_{\rm BF}$ as
\begin{align}
&\Delta_{\rm FS} \equiv \beta \bigl\{F(F_{\rm min})-F(S_{\rm min})\bigr\},\\
&\Delta_{\rm BF} \equiv \beta \bigl\{F(B)-F(F_{\rm min})\bigr\},
\end{align}
where $F(F_{\rm min})$, $F(S_{\rm min})$, and $F(B)$ are the free energies at $F_{\rm min}$, $S_{\rm min}$, and $B$, 
respectively (see Fig.~\ref{fig:Def_DF}). 
We measure $\Delta_{\rm FS}$ and $\Delta_{\rm BF}$ at temperatures below $T_{\rm cross}$ 
and observe how they changes as the temperature decreases from $T_{\rm cross}$. 
In concrete, $\Delta_{\rm FS}$ and $\Delta_{\rm BF}$ are evaluated for each sample at temperatures
\begin{align}
T=T_{\rm cross}-0.1nJ\quad(n=0,1,2,\cdots),
\label{eqn:def_Tana}
\end{align}
by interpolation. This measurement is done above the minimum temperature $T_{\rm min}=1.0~J$. 

\begin{figure}[h]
\begin{center}
\includegraphics[width=7cm]{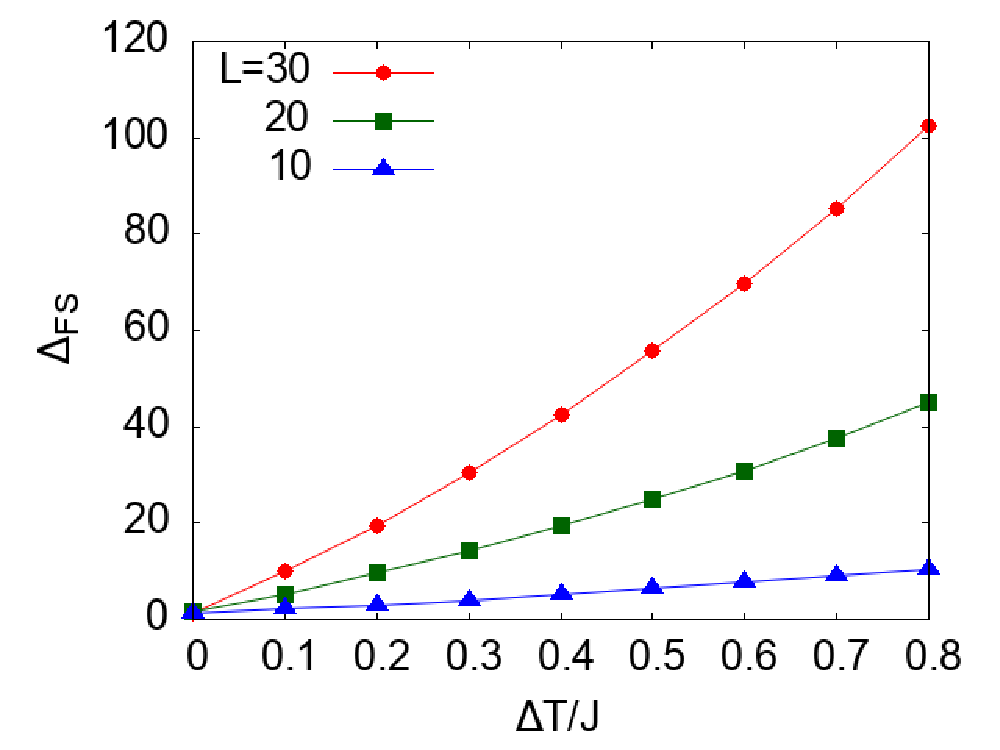}
\end{center}
\caption{The free-energy difference $\Delta_{\rm FS}$ averaged over samples is plotted as a function of $\Delta T$, 
where $\Delta T$ is the difference between the measurement temperatures given by Eq.~\eqref{eqn:def_Tana} 
and $T_{\rm cross}$. The coefficient $C^{\rm F}$ is $0.90$.}
\label{fig:DeltaFS}
\end{figure}

In Fig.~\ref{fig:DeltaFS}, $\Delta_{\rm FS}$ averaged over samples is plotted as a function of $\Delta T$, where 
$\Delta T=0.1nJ$ is the difference between the measurement temperatures 
given by Eq.~\eqref{eqn:def_Tana} and $T_{\rm cross}$. 
Because $T_{\rm cross}$ and the maximum value of $\Delta T$ that satisfies the inequality 
$T_{\rm cross}-\Delta T \ge T_{\rm min}$ 
are different from sample to sample, the number of samples used for the average depends on $\Delta T$. 
When $\Delta T$ is $0$, $F(F_{\rm min})$ is almost the same as $F(S_{\rm min})$ 
because the temperature is just $T_{\rm cross}$. 
Therefore, $\Delta_{\rm FS}$ at $\Delta T=0$ is almost zero for all $L$. 
We see that $\Delta_{\rm FS}$ increases with increasing $\Delta T$, 
and its speed increases with increasing $L$. 
This means that $S_{\rm min}$ is stabilized with decreasing temperature, 
and the stabilization becomes faster with increasing $L$. 

\begin{figure}[h]
\begin{center}
\includegraphics[width=7cm]{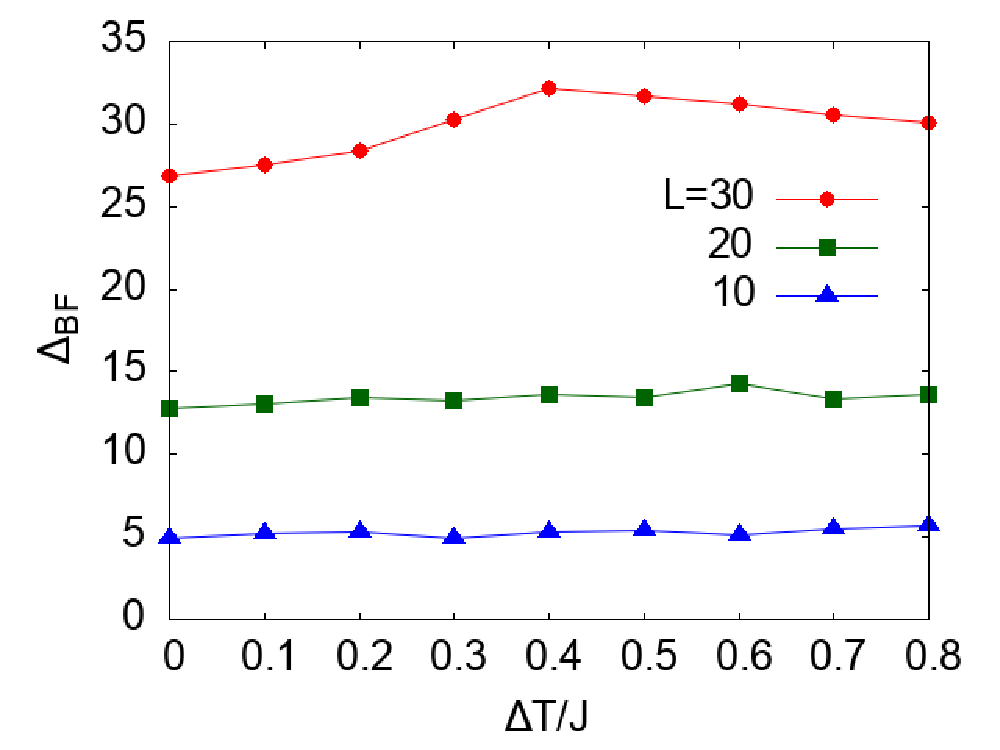}
\end{center}
\caption{The free-energy difference $\Delta_{\rm BF}$ averaged over samples is plotted as a function of $\Delta T$, 
where $\Delta T$ is the difference between the measurement temperatures given by Eq.~\eqref{eqn:def_Tana} 
and $T_{\rm cross}$. The coefficient $C^{\rm F}$ is $0.90$.}
\label{fig:DeltaBF}
\end{figure}

We next show the $\Delta T$ dependence of $\Delta_{\rm BF}$ in Fig.~\ref{fig:DeltaBF}. 
Interestingly, $\Delta T$ dependence of $\Delta_{\rm BF}$ is rather weak. 
We also notice that $\Delta_{\rm BF}$ increases with increasing $L$. 
When $L=30$ and $\Delta T=0$, $\Delta_{\rm BF}$ is about $26.9$. 
This means that it is quite difficult for the system to escape from the local minimum $F_{\rm min}$ 
because the probability that the system overcomes the barrier $B$ is given by $\exp(-\Delta_{\rm BF})$. 

\begin{figure}[h]
\begin{center}
\includegraphics[width=7cm]{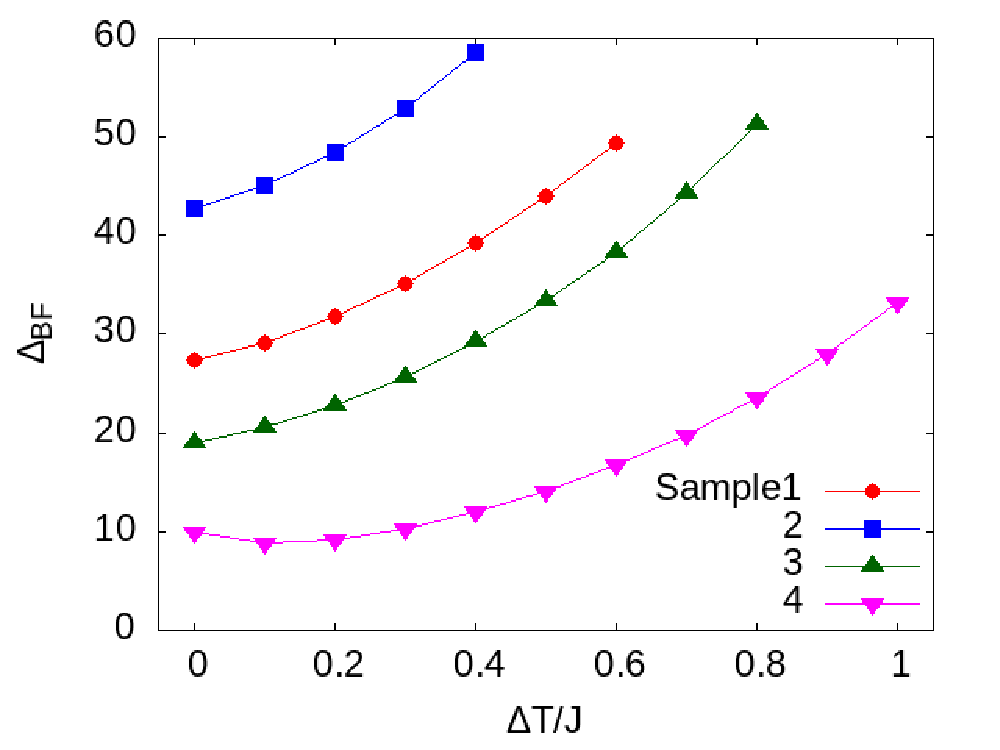}
\end{center}
\caption{The temperature dependence of $\Delta_{\rm BF}$ for four samples. The size $L$ is $30$ and the 
coefficient $C^{\rm F}$ is $0.90$. The samples are the same as those in Fig.~\ref{fig:SampMag_L030_CMF090}.}
\label{fig:SampDeltaBF_L030_CMF090}
\end{figure}

In Fig.~\ref{fig:SampDeltaBF_L030_CMF090}, $\Delta_{\rm BF}$'s for four samples are plotted 
as a function of $\Delta T$. The size $L$ is $30$ and the samples are the same as those in Fig.~\ref{fig:SampMag_L030_CMF090}. 
We see that the behavior of $\Delta_{\rm BF}$ varies considerably depending on the sample.

\subsection{Annealing simulation}
\begin{figure}[h]
\begin{center}
\includegraphics[width=7cm]{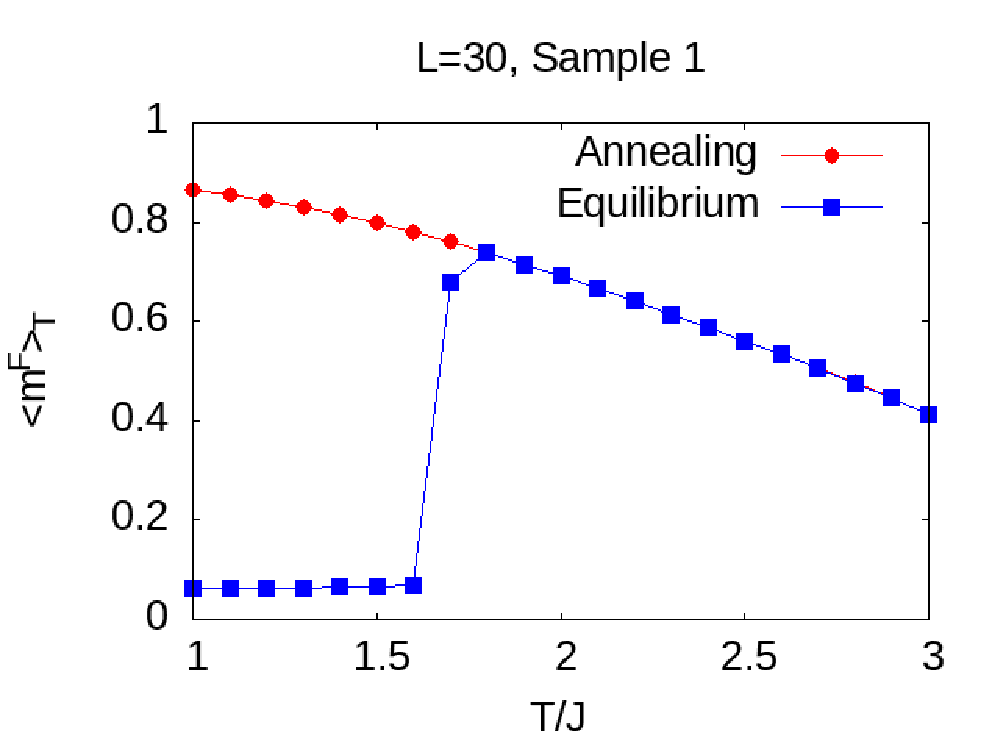}
\end{center}
\caption{The temperature dependences of $\langle m^{\rm F} \rangle_T$ obtained by 
annealing simulation (full circles) and equilibrium simulation (full squares). 
The size $L$ is $30$ and the coefficient $C^{\rm F}$ is $0.90$. 
The sample is the same as ”Sample 1” in Figs.~\ref{fig:SampMag_L030_CMF090} 
and \ref{fig:SampDeltaBF_L030_CMF090}.}
\label{fig:Annealing_L030_CMF090}
\end{figure}

The result shown in Fig.~\ref{fig:DeltaBF} indicates that, if the size $L$ is large and the free-energy barrier 
$\Delta_{\rm BF}$ is high, the system cannot escape from $F_{\rm min}$ even when the temperature 
becomes lower than $T_{\rm cross}$ and $F_{\rm min}$ becomes globally unstable. To verify the validity of this 
conjecture, we performed an annealing simulation. The result is shown in Fig.~\ref{fig:Annealing_L030_CMF090}.  
In this simulation, the system was gradually cooled from an initial temperature $6.0~J$ to $1.0~J$ 
in steps of $\Delta T=0.1~J$. At each temperature, a usual MC simulation with the Metropolis method~\cite{Metropolis53} was performed. 
The initial temperature was set to be well above the critical temperature 
$T_{\rm c}^{\rm F}$. The system was kept at each temperature for $2\times 10^7$ MCS. 
The first $10^7$ MCS are for equilibration and the following $10^7$ MCS are for measurement. 

In Fig.~\ref{fig:Annealing_L030_CMF090}, $\langle m^{\rm F} \rangle_T$ is plotted as a function of temperature. 
The result of the annealing simulation is denoted by full circles. The full squares denotes 
equilibrium data obtained by the variant of the Wang-Landau method. The size $L$ is $30$ and 
the coefficient $C^{\rm F}$ is $0.90$. The sample is the same as ”Sample 1” in Figs.~\ref{fig:SampMag_L030_CMF090} 
and \ref{fig:SampDeltaBF_L030_CMF090}. Therefore, the data shown in Fig.~\ref{fig:SampMag_L030_CMF090} and 
the equilibrium data shown in Fig.~\ref{fig:Annealing_L030_CMF090} are the same. 
Although the annealing and equilibrium data coincide with each other at high temperatures, they do not at low temperatures. 
This means that the system was not equilibrated in the annealing simulation at low temperatures. 
We performed annealing simulation $10$ times with different random number sequences, but the results were almost the same. 
These results indicate that, as conjectured, the system continues to remain at $F_{\rm min}$ at low temperatures 
despite that $F_{\rm min}$ becomes globally unstable. It is worth pointing out that these results are consistent with 
the data of $\Delta_{\rm BF}$ shown in Fig.~\ref{fig:SampDeltaBF_L030_CMF090} (full circles), which shows that 
$\Delta_{\rm BF}$ is about $27.3$ at $\Delta T=0$ and it increases with increasing $\Delta T$. 
Because the probability that the system overcomes the barrier is given by $\exp(-\Delta_{\rm BF})$, 
the probability is about $1.4\times 10^{-12}$ at $T_{\rm cross}$ ($\Delta T=0$) and it decreases with decreasing temperature. 
Therefore, it is quite difficult for the system to overcome the barrier within $2\times 10^7$ MCS, i.e., 
the period at each temperature in this annealing simulation.

\section{Summary and Discussion}
\label{sec:Summary_Discussion}
In this paper, we presented a simple model that recalls two different patterns depending on the temperature. 
To realize a change in recall pattern due to temperature change, we embed two patterns to different graphs: 
the first pattern into a fully connected graph and the second pattern into a sparse graph. 
Because a fully connected graph is more resistant to thermal fluctuations than a sparse graph, 
we can realize a change in recall pattern by tuning relative weights of the two patterns properly. 
In this study, we chose the two dimensional square lattice as a sparse graph. We demonstrated 
by equilibrium Monte-Carlo simulations that such a temperature-dependent change in recall patterns does occur in our model. 
Simulation results strongly indicate that the system undergoes a first-order phase transition when the change in recall patterns occurs. 

We also found that the free-energy barrier $\Delta_{\rm BF}$ increases with the system size. 
Therefore, as the size increases, the system has difficulty recalling patterns embedded in the sparse graphs at low temperatures 
because it becomes difficult for the system to overcome the the free-energy barrier within the given simulation timescale 
(see Fig.~\ref{fig:Annealing_L030_CMF090}). 
It is a future challenge to modify the model so as to suppress the free-energy barrier for a smooth switching of recall patterns. 

In this study, only two patterns were embedded in the model. Therefore, the recall pattern was switched only once due to the temperature change.
However, by preparing many graphs with different resistance to thermal fluctuations and embedding different patterns in each graph, 
it is possible to modify the model so that the recall pattern is switched multiple times. 
To prepare graphs with different resistance to thermal fluctuations, one may use the geometric graph~\cite{Sasaki20} in which 
vertices located on a $d$-dimensional hypercubic lattice are connected by an edge with a probability proportional to $d_{ij}^{-\alpha}$, 
where $d_{ij}$ is the distance between the two vertices and $\alpha$ is a positive exponent. 
By changing the exponent $\alpha$, we can continuously change the effective dimension of the graph, 
and therefore its resistance to thermal fluctuations. On the other hand, an increase in the number of embedded patterns 
is expected to obscure the switching of recall patterns with the temperature change. 
This is the reason why we restrict ourselves to the case that the number of embedded patterns is two.


%





\appendix 
\section{Proof for the existence of sequential transitions}
\label{sec:proof}
Sequential transitions mentioned in Sect.~\ref{sec:Model} occur 
if we choose $C^{\rm F}$ so as to satisfy the following two conditions:
\begin{itemize}
\item $T_{\rm c}^{\rm S} <  T_{\rm c}^{\rm F}$.
\item $E_{\rm GS}^{\rm S} < E_{\rm GS}^{\rm F}$.
\end{itemize}
From Eqs.~\eqref{eqn:Tc_F} and \eqref{eqn:E_GS}, 
we find that there exists $C^{\rm F}$ that satisfies both the two conditions if the inequality
\begin{equation}
T_{\rm c}^{\rm S} < kJ,
\label{eqn:condition}
\end{equation}
holds. The transition temperature $T_{\rm c}^{\rm S}$ is maximized 
when the graph is a random one because the effective spatial dimension of the random graph is infinity. 
The transition temperature $T_{\rm c}^*(k)$ for random graphs with a fixed degree $k$ is 
analytically calculated~\cite{MezardMontanari09} as 
\begin{equation}
T_{\rm c}^*(k)=\frac{J}{\tanh^{-1}\left(\frac{1}{k-1}\right)}.
\end{equation}
Because $T_{\rm c}^*(k) < kJ$ for $k \ge 3$ and $T_{\rm c}^{\rm S} \le T_{\rm c}^*(k)$, 
the inequality \eqref{eqn:condition} is always satisfied. Therefore, it is proven that 
we can always choose $C^{\rm F}$ for any sparse graphs so that sequential transitions occur.

\end{document}